\documentclass[%
reprint,
twocolumn,
superscriptaddress,
showpacs,preprintnumbers,
nofootinbib,
aps,
prl
]{revtex4-2}

\usepackage{epsfig}
\usepackage{amsfonts}
\usepackage{bbm,bm}
\usepackage{comment}
\usepackage{amsmath}

\usepackage[breaklinks,urlbordercolor={1 1 1}]{hyperref}
\usepackage[utf8]{inputenc}
\usepackage{graphicx}
\usepackage[dvips]{xcolor}
\graphicspath{ {./figures/} }
\let\vec\mathbf 

\newcommand{\nn}{\nonumber \\}
\newcommand{\rmi}[1]{{\mbox{\scriptsize #1}}}
\newcommand{\rmii}[1]{{\mbox{\tiny\rm{#1}}}}

\newcommand{\mH}{m_\rmii{$H$}}

\newcommand{\Tc}{T_{\rm c}}

\newcommand{\Tint}[1]{{\hbox{$\sum$}\!\!\!\!\!\!\!\int\,}_{\!\!\!\!\raise-0.9ex\hbox{$\scriptstyle{#1}$}}}
\newcommand{\Tinti}[1]{{{\Sigma}\!\!\!\!\raise0.3ex\hbox{$\int$}_\rmii{${#1}$}}}
\newcommand{\Tintip}[1]{{{\Sigma'}\!\!\!\!\!\raise0.3ex\hbox{$\int$}_\rmii{${#1}$}}}

\newcommand{\Veff}{V^{\rmi{eff}}}
\newcommand{\Seff}{S^{\rmi{eff}}}

\newcommand{\define }{\equiv}

\newcommand{\diff}{{\rm d}}
\newcommand{\ordo}[1]{\mathcal{O}({#1})}

\renewcommand{\vec}[1]{{\bf #1}}

\begin{document}
\newcommand{\UPP}{\affiliation{
Department of Physics and Astronomy, Uppsala University,
Box 516, SE-751 20 Uppsala,
Sweden}}

\newcommand{\HEL}{\affiliation{%
Department of Physics and Helsinki Institute of Physics,
PL 64, FI-00014 University of Helsinki,
Finland}}

\newcommand{\NOR}{\affiliation{Nordita,
KTH Royal Institute of Technology and Stockholm University,
Roslagstullsbacken 23,
SE-106 91 Stockholm,
Sweden}}

\newcommand{\TSUa}{\affiliation{Tsung-Dao Lee Institute and School of Physics and Astronomy,
Shanghai Jiao Tong University,
800 Dongchuan Road, Shanghai, 200240 China}}

\newcommand{\TSUb}{\affiliation{Shanghai Key Laboratory for Particle Physics and Cosmology,
Key Laboratory for Particle Astrophysics \& Cosmology (MOE),
Shanghai Jiao Tong University,
Shanghai 200240, China}}

\newcommand{\CAL}{\affiliation{Kellogg Radiation Laboratory, California Institute of Technology,
Pasadena,
CA~91125 USA}}

\newcommand{\UMA}{\affiliation{Amherst Center for Fundamental Interactions, Department of Physics,
University of Massachusetts, Amherst,
MA~01003, USA}}

\title{Nucleation at finite temperature: a gauge-invariant, perturbative framework}

\preprint{ACFI-T21-15}
\preprint{HIP-2021-44/TH}
\preprint{NORDITA 2021-110}

\author{Johan L{\"o}fgren}
\email{johan.lofgren@physics.uu.se}
\UPP

\author{Michael J.~Ramsey-Musolf}
\email{mjrm@sjtu.edu.cn, mjrm@physics.umass.edu}
\UMA \CAL \TSUa \TSUb

\author{Philipp Schicho}
\email{philipp.schicho@helsinki.fi}
\HEL

\author{Tuomas V.~I.~Tenkanen}
\email{tuomas.tenkanen@su.se}
\TSUa \TSUb \NOR

\begin{abstract}
We present a gauge-invariant framework for bubble nucleation in
theories with radiative symmetry breaking at high temperature.
As
a procedure,
this perturbative framework establishes a practical, gauge-invariant computation of 
the leading order nucleation rate,
based on a consistent power counting in the high-temperature expansion.
In model building and particle phenomenology, this framework has applications such as
the computation of
the bubble nucleation temperature and
the rate for electroweak baryogenesis and
gravitational wave signals from cosmic phase transitions.
\end{abstract}

\maketitle

Achieving a rigorous understanding of the thermal history of the electroweak (EW) symmetry breaking has been a long-standing challenge at the interface between 
particle physics and cosmology. 
The standard history of the Standard Model (SM) of particle physics contains
no electroweak phase transition. 
Instead a smooth crossover between high and low temperature phases occurs~\cite{Kajantie:1996mn,Csikor:1998eu}. 
A non-standard history with a first order phase transition
is conceivable in theories beyond the SM (BSM) that are well motivated from
the electroweak to TeV scale.
BSM scenarios with extended scalar sectors can contain rich patterns of
symmetry breaking~\cite{Weinberg:1974hy} 
and potentially produce 
EW baryogenesis~\cite{Kuzmin:1985mm,Shaposhnikov:1987tw,Morrissey:2012db}, and also
rich phenomenology for high energy collider physics \cite{Ramsey-Musolf:2019lsf} 
and production of primordial gravitational waves~\cite{Caprini:2019egz}.
These phenomena have been actively
studied for decades and
recently sparked increased interest in
next-generation gravitational wave detector
experiments~\cite{Kawamura:2011zz,Harry_2006,Ruan:2018tsw}
such as
LISA~\cite{Audley:2017drz}. 

One key ingredient for both EW baryogenesis and
stochastic gravitational wave background production is
a first order phase transition in the primordial plasma of particles that
proceeds via nucleating bubbles of the low-temperature phase.
The thermo- and bubble dynamics of such transitions can be 
reliably described non-perturbatively using
lattice simulations~\cite{Farakos:1994xh,Kajantie:1995kf,Moore:2000jw}.
However, exploring BSM scenarios and their thermodynamics invariably requires
the use of perturbation theory,
since fully comprehensive non-perturbative analyses of
a multidimensional parameter space are computationally out of reach.
A qualitative and quantitative assessment of the perturbative reliability
requires robust theory computations when comparing to lattice data. 
In that respect,
an unphysical gauge-dependence of the bubble nucleation rate has plagued 
particle physics phenomenology applications of
the thermal phase transition literature for decades 
(c.f.~\cite{Caprini:2015zlo,Caprini:2019egz} and references therein).
While by now
resolved
at zero temperature~\cite{%
  Metaxas:1995ab,Baacke:1999sc,Andreassen:2014eha,Andreassen:2014gha,
  Garbrecht:2015yza,Endo:2017gal},
a similar, long-standing and open problem exists
at finite temperature
if the potential barrier between the phases is radiatively generated.

We resolve this long-standing problem
by providing a practical, gauge-invariant framework for
thermal bubble nucleation, intended
for model-building and particle phenomenology applications. 
The reliability of the approach can be assessed by ``benchmarking'', against
lattice analyses in the future.
Such benchmarking is incompatible with the conventional approach since there
the nucleation rate is gauge-dependent and
an ill-defined unphysical quantity.

The bubble nucleation rate per unit volume, 
$\Gamma$, 
has the semiclassical approximation~\cite{Langer:1969bc,Coleman:1977py,Linde:1980ts,Affleck:1980ac}
\begin{align}
\label{eq:rate}
  \Gamma &= A e^{-\mathcal{B}}
  \;.    
\end{align}
Here,
the exponent $\mathcal{B}=S_3/T$ with $S_3$ being the three-dimensional Euclidean effective action evaluated at the ``bounce'' solution that
solves the classical Euclidean field equations~\cite{Coleman:1977py,Linde:1981zj}.
The prefactor $A$ is dimensionful and given by
the characteristic mass scales of the theory, 
resulting from computing the functional determinants.
The leading behaviour of the rate $\Gamma$ is encoded in the exponent $\mathcal{B}$.

The effective action is computable using the background field ($\phi$) method.
Therein,
$\Veff(\phi,T)$
is the thermal effective potential describing the equilibrium free energy of the system
and
$Z^{\frac{1}{2}}(\phi,T)$
is the field renormalization factor. 
Both $\Veff(\phi,T)$ and  $Z(\phi,T)$ admit an expansion in the weak gauge coupling, denoted generically here as $g$.
After the gradient expansion in powers of spatial derivatives $(\partial_i \phi)$,
the effective action reads \cite{Langer:1974xx,Moss:1985ve}
\begin{equation}
\label{eq:action3}
S_3= \int {\rm d}^3 x \Bigl[
    \Veff(\phi,T)
  + \frac{1}{2}Z(\phi,T)\left(\partial_i \phi\right)^2
  + \dots
\Bigr]
  \;,
\end{equation}
where the
ellipsis
contains terms of additional $(\partial_i \phi)$-powers.
While
the gradient expansion in general does not converge~\cite{%
  Moore:2000jw,Garny:2012cg,Gould:2021ccf,Hirvonen:2021xxx},
one may obtain
a self-consistent, gauge-invariant estimate of $\Gamma$ by
ensuring the two following conditions apply
\begin{itemize}
\item[(A):]
  Fields that generate the barrier must be parametrically heavier than
  the nucleating field.
  In this case leading orders can be correctly described by a derivative expansion, 
\item[(B):]
  Considering temperatures in the vicinity of the critical temperature, $\Tc$,
  such that a specific power counting in $g$ applies,
  wherein the leading order potential has a radiatively generated barrier. 
\end{itemize}
Hence, we focus on a temperature regime where 
the first two leading terms of the effective action are enhanced by powers of $g$ 
and
presentable schematically as 
\begin{align}
\label{eq:Seff-expansion}
S_3 &=
    a_0 g^{-\frac{3}{2}}
  + a_1 g^{-\frac{1}{2}}
  + \Delta
\;. 
\end{align}
Here $a_{0,1}$ are numerical coefficients that are computable in 
the gradient expansion at
leading (LO) and
next-to-leading orders (NLO), respectively. 
$\Delta$ encodes formally higher order corrections 
--
that are $\ordo{1}$, up to potential logarithms of $g$ \cite{Gould:2021ccf}
--
that we presently do not compute. 
The bubble nucleation rate then reads
\begin{align}
\label{eq:rate-LO}
  \Gamma &= A' e^{-
    \left(
        a_0 g^{-\frac{3}{2}}
      + a_1 g^{-\frac{1}{2}}
      \right)
    }  
\;.
\end{align}
We formally collected higher order effects encoded in
$\Delta$ in Eq.~\eqref{eq:Seff-expansion}, to
a yet unspecified prefactor $A' \equiv A e^{-\Delta}$.
The expression~\eqref{eq:rate-LO} results from an expansion based on
the chain of scale hierarchies at high temperature
\begin{align}
\label{eq:thermal-hierarcy}
\pi T \gg g T \gg g^{\frac{3}{2}} T
\;,
\end{align}
where
$\pi T$ is the thermal scale of non-zero Matsubara modes,
$g T$ is the intermediate scale of zero modes that are thermally screened,
and
$g^{\frac{3}{2}} T$ is a characteristic scale of bubble nucleation, related to an effective mass of the nucleating field.
For a systematic treatment of thermodynamics in the framework of three-dimensional, high temperature effective field theory (3d EFT)
see~\cite{Kajantie:1995dw,Braaten:1995cm}.
This framework was recently extended~\cite{Gould:2021ccf}
by presenting an approach for an effective description for bubble nucleation. 

Here,
we work at
the leading high temperature expansion~\cite{Arnold:1992rz} that allows us to
perform the computation without formally constructing a 3d EFT.   
Our goal is to
compute LO and NLO terms in 
$\mathcal{B}$ 
and
show their gauge invariance.
We emphasize that this computation agrees with
the generic EFT approach~\cite{Gould:2021ccf} at leading orders, and
an accompanying article~\cite{Hirvonen:2021xxx} revisits this computation in
the EFT context, which allows to pursue improvement in terms of higher order corrections.

As in earlier literature both at
zero~\cite{Metaxas:1995ab,Arunasalam:2021zrs} and
high~\cite{Garny:2012cg} temperature,
we use
the Abelian Higgs model as a concrete illustration. 
The Euclidean Lagrangian density for the Abelian Higgs model is
\begin{align}
\label{eq:lag4d}
\mathcal{L}_{} &=
    \frac{1}{4} F_{\mu\nu}F_{\mu\nu}
  + (D_\mu \Phi)^* (D_\mu \Phi)
  \nn &
  + \mu^2 \Phi^* \Phi
  + \lambda (\Phi^* \Phi)^2 \nonumber 
  + \mathcal{L}^{ }_{\rmii{GF}}
  + \mathcal{L}^{ }_{\rmii{FP}} 
  \;,
\end{align}
with
a U(1) gauge field $B_\mu$ and
a complex scalar field $\Phi$.
The covariant derivative for complex Higgs reads
$D_\mu \Phi = \partial_\mu \Phi - ig B_\mu \Phi$, 
where $g$ is the gauge coupling
and
the field strength tensor
$F_{\mu\nu} =\partial_\mu B_\nu-\partial_\nu B_\mu$.
We expand $\Phi$ in real fields 
\begin{equation}
\label{eq:ComplexPhi}
  \Phi= \frac{\phi}{\sqrt{2}} + \frac{1}{\sqrt{2}}\bigl(H+i \chi\bigr)
  \;,
\end{equation}
where
$\phi \equiv \phi(x)$ is a spatially dependent classical background field and
$H$ and $\chi$ are propagating quantum degrees of freedom.
In perturbation theory
the gauge can be fixed using
$R_\xi$-gauge~\cite{Fukuda:1975di,Garny:2012cg} 
\begin{align}
\label{eq:Rxi:F}
\mathcal{L}^{}_{\rmii{GF}} &=
  \frac{1}{2\xi}  \bigl[ -\bigl(\partial_\mu B_\mu
  + i g \xi (\tilde{\phi}^* \Phi - \Phi^* \tilde{\phi})
  \bigr) \bigr]^2
  \;,
\end{align}
with Fadeev-Popov ghost ($c,\bar{c}$) Lagrangian
\begin{align}
\label{eq:L:FP}
\mathcal{L}_{\rmii{FP}} &=
\bar c \Big( -\square + \xi g^2 (\tilde{\phi}^* \Phi + \Phi^* \tilde{\phi}) \Big)c
\;.
\end{align}
Although one need not relate $\tilde{\phi}$  directly to $\phi$, it is convenient
to identify
$\tilde{\phi} = \phi/\sqrt{2}$
to eliminate mixing between the gauge field and Goldstone mode $\chi$. 

Equilibrium thermodynamics are described by the effective potential, whose
 minima should be separated by a barrier for a first order phase transition.
While the Abelian Higgs model admits no tree-level barrier,
the barrier arises radiatively through loop corrections. 
Similar to~\cite{Metaxas:1995ab,Garny:2012cg,Arnold:1992rz,Ekstedt:2020abj},
we show that
one-loop gauge field contributions can arise at
the same order as
the tree-level potential provided that model parameters assume 
the following power counting.
First, we assume that
\begin{align}
\label{eq:phi-scaling}
\phi &\sim T
\;,
\end{align}
which sets the size of the background field to the characteristic mass scale of
the problem, the temperature. 
Second, we relate the size of the scalar self-coupling and gauge coupling
by~\cite{Arnold:1992rz}
\begin{align}
\label{eq:lam-scaling}
\lambda &\sim g^3
\;.
\end{align}
By assuming that
the quartic coupling is sufficiently small compared to gauge coupling,
the cubic term induced by the gauge field will contribute at leading order, as seen below.
Third, we assume
\begin{align}
\label{eq:mu-scaling}
\mu^2 &\sim (g T)^2
\;,
\end{align}
which installs the high-temperature expansion of $\mu^2 \ll T^2$.
Finally, the effective thermally corrected mass of the scalar at leading order is 
\begin{align}
\label{eq:mu-eff-scaling}
\mu^2_{\rmii{eff}} \define \mu^2
  + \bigl(4\lambda+3g^2\bigr)\frac{T^2}{12} &\sim
    \ordo{g^{2+N} T^2}
 \;.
\end{align}
Here,
the first and third terms are individually of $\ordo{g^2 T^2}$.
However, for $\mu^2<0$, a cancellation between the first and remaining terms can render 
the overall sum in $\mu_{\rmii{eff}}^{2}$ smaller -- parametrised by $N>0$ --
for some temperature range.
Henceforth, we
assume a temperature window where $N=1$ and
argue below that this occurs in the vicinity of
the phase transition critical temperature \cite{Arnold:1992rz,Gynther:2005av}.

The assumptions of Eqs.~\eqref{eq:phi-scaling}--\eqref{eq:mu-eff-scaling}
induce the 
chain of thermal scale hierarchies in
Eq.~\eqref{eq:thermal-hierarcy}.
Non-zero Matsubara modes have masses $\sim\pi T$.
In the unbroken phase ($\phi = 0$)
zero Matsubara modes 
have the following scales:
Higgs and Goldstone fields have masses
$\sim \mu_{\rmii{eff}} \sim g^{\frac{3}{2}}T$.
At zero temperature it is sufficient to talk of transverse and longitudinal modes of the gauge field, but at high temperatures there is a further splitting of the three transverse modes into two spatial modes and one temporal.
The gauge field temporal mode, $B_0$, has
a mass $\bar{m}_{\rmii{$B_0$}}\sim g T$;
while spatial 
gauge fields remain massless ($m_{\rmii{$B$}}=0$).
In the broken phase ($\phi > 0$),
spatial and temporal 
gauge bosons, Goldstone and ghost fields have
masses $\sim g \phi \sim g T$, while 
the Higgs field that undergoes nucleation has parametrically lighter mass
$\sim g^{\frac{3}{2}} T$.

The full one-loop effective potential is found using
the following background field dependent mass eigenvalues
\begin{align}
\label{eq:mh3sq}
\bar{m}^2_{\rmii{$H$}} &=
  \mu^2_{\rmii{eff}} + 3 \lambda \phi^2
  \;, \\[2mm]
\bar{m}^2_{\chi} &=
  \mu^2_{\rmii{eff}} + \lambda \phi^2 + g^2 \xi \phi^2
  \;, \\
\label{eq:mB0sq}
\bar{m}_{\rmii{$B_0$}}^2 &=
  \frac{1}{3} g^2 T^2 + g^2 \phi^{2}
  \;, \\
  m^2_{c} &=
  g^2 \xi \phi^2
  \;, \quad 
  m_{\rmii{$B$}}^2 =
  g^2 \phi^2
  \;,
\end{align}
where masses for the zero Matsubara modes of the
Higgs,
Goldstone 
and temporal gauge field $B_0$ include resummed
thermal corrections
e.g.\ the $T^2$-term in the third line describes Debye screening.
The mass of the longitudinal component of the gauge field equals the ghost mass.
Neither spatial
gauge fields, nor ghost fields, 
develop thermal masses.
Based on the power counting in $g$, the leading contribution to the effective potential is of $\ordo{g^3 T^4}$ and reads 
\begin{align}
\label{eq:VLO}
\Veff_{\rmii{LO}} &= 
\frac{1}{2} \mu^2_{\rmii{eff}} \phi^2 + \frac{1}{4} \lambda \phi^4
  \nn &
  - \frac{g^3 T}{12\pi} \Bigl[
      2 \phi^3
    + \Bigl(\frac{1}{3}T^2 + \phi^2\Bigr)^{\frac{3}{2}}
  \Bigr]
\;,
\end{align}
where
the second line is the transverse gauge field contribution and
the second term therein corresponds to the Debye mass~\eqref{eq:mB0sq} of the temporal mode. 
Near the phase transition, all terms in
the potential should be 
approximately 
of the same order of magnitude which is
assured by construction,
given the assumed power counting in $g$
in Eqs.~\eqref{eq:phi-scaling}--\eqref{eq:mu-eff-scaling}.
The leading-order potential of Eq.~\eqref{eq:VLO} at $\ordo{g^3 T^4}$ is gauge invariant.

The one-loop contribution to the effective potential from
the longitudinal gauge field, Goldstone field, and ghost fields is
\begin{align}
\label{eq:goldstone-ghost}
  - \frac{T}{12\pi} \left( \bar{m}_\chi^3 - m_{c}^3  \right) \sim
  \ordo{g^4 T^4}
  \;.
\end{align}
Note that ghosts contribute with a relative minus sign to other fields. 
These contributions give rise to
an explicit $\xi$ dependence,
but
these terms are of higher order compared to Eq.~\eqref{eq:VLO} due to
a cancellation at leading order.
Below we include the remaining $\ordo{g^4 T^4}$ terms at NLO in the effective potential 
which expands as
\begin{align}
\label{eq:expandingThermalPotential}
  \Veff &=
      \Veff_{\rmii{LO}}
    + \Veff_{\rmii{NLO}}
    + \ordo{g^{\frac{9}{2}} T^4}
    \;,
\end{align}
where
\begin{align}
\label{eq:VNLO}
\Veff_{\rmii{NLO}} &= \frac{1}{(4\pi)^2} \bigg\{ 
  g^4 T^2 \phi^2 \bigg(-1 + \ln \Big( \frac{4g^2 \phi^2}{\Lambda^2} \Big) \bigg)
  \nn &
  + \sqrt{\xi} g T \phi \Big( g^3 T \phi - 2\pi (\mu^2_{\rmii{eff}} + \lambda \phi^2) \Big) 
  \nn &
  + g^4 T^2 \bigg[ \frac{1}{2} \sqrt{\xi} \phi \sqrt{\frac{1}{3}T^2 + \phi^2}
  \nn &
  + \frac{1}{2} \phi^2 \bigg(-1 + \ln \Big( \frac{4 g^2 (\frac{1}{3}T^2 + \phi^2)}{\Lambda^2} \Big) \bigg)
\bigg]
\bigg\}
\;.
\end{align}
The last term on the second line originates from
the one-loop Goldstone-ghost contribution in Eq.~\eqref{eq:goldstone-ghost}, and
the last two lines correspond to
the two-loop contributions of the
$B_0$ field.
The remaining terms result from two-loop diagrams with
spatial gauge fields, ghosts and scalars ($\chi$,$H$).
The $\ordo{g^{\frac{9}{2}} T^4}$ term in Eq.~\eqref{eq:expandingThermalPotential}
arises at one-loop order from the Higgs field.

Consistent treatment of the effective action~\eqref{eq:action3} 
at NLO
also requires the inclusion of field renormalization: 
\begin{align}
\label{eq:Z:exp}
  Z &= 1
    + Z_{\rmii{NLO}}
    + \ordo{g^{\frac{3}{2}}}
    \;,
\end{align}
where
\begin{equation}
\label{eq:Z-NLO}
  Z_{\rmii{NLO}}(\phi)= \frac{gT}{48 \pi} \biggl[
    - \frac{22}{\phi}
    + \frac{\phi^2}{(\frac{1}{3} T^2 + \phi^2)^{\frac{3}{2}}}
  \biggr]
  \;.
\end{equation}
The first term is from the two spatial modes and ghosts and the second term is from the temporal mode. 
Notably, at order $\ordo{g}$
the field renormalisation is independent of the gauge-fixing parameter.
Higher order terms at $\ordo{g^{\frac{3}{2}}}$ arise from two-loop diagrams involving
gauge, ghost and Goldstone fields, and
one-loop diagrams with internal Higgs legs.

In the semiclassical approximation,
the background field extremizes the leading-order action and can be found from the equation of motion for the leading-order potential  
\begin{equation}
\label{eq:thermalLOeom}
  \nabla^2 \phi_b(x) = \frac{\partial \Veff_{\rmii{LO}}} {\partial\phi}\biggr|_{\phi = \phi_b}
  \;,\quad
  \begin{cases}
    \phi_b(\infty) &= 0 \\
    \phi_b'(0) &= 0
  \end{cases}
  \;,
\end{equation}
where 
$\nabla^2 \equiv\partial_i\partial_i$
is the three-dimensional Laplacian operator,
and
$\phi_b$ is 
the ``bounce solution''~\cite{Coleman:1973jx}.
We expand the exponent of the nucleation rate as
$\mathcal{B} = \mathcal{B}_0 + \mathcal{B}_1$:
\begin{align}
\label{eq:B0}
\mathcal{B}_0 &=
  \beta \int \diff^3 x \left[
      \Veff_{\rmii{LO}}(\phi_b)
    + \frac{1}{2}\left(\partial_i \phi_b\right)^2
  \right]
  \;,\\
\label{eq:B1}
\mathcal{B}_1 &=
  \beta \int \diff^3 x \left[
      \Veff_{\rmii{NLO}}(\phi_b)
    + \frac{1}{2}Z_{\rmii{NLO}}^{ }(\phi_b)\left(\partial_i \phi_b\right)^2
  \right]
  \;, 
\end{align}
where $\beta \equiv 1/T$. 
The characteristic length scale $R$ for nucleation is related to
the typical bubble size, given by
the inverse mass of the nucleating field
$R \sim \mH^{-1} \sim g^{-\frac{3}{2}} T^{-1}$.
This gives rise to the formal scaling
$\int\diff{}^3x \sim g^{-\frac{9}{2}} T^{-3}$.
Together with the power counting for the effective potential and field renormalisation,
this establishes the relative importance of the first two leading exponent terms of
the nucleation rate:
$\mathcal{B}_0 \sim g^{-\frac{3}{2}}$,
$\mathcal{B}_1 \sim g^{-\frac{1}{2}}$
as already foreshadowed in Eqs.~\eqref{eq:Seff-expansion} and \eqref{eq:rate-LO}.
Despite the
$1/\phi$ behaviour of $Z_{\rmii{NLO}}^{ }(\phi_b)$,
the contribution of the term
$Z_{\rmii{NLO}}^{ }(\phi_b)\left(\partial_i \phi_b\right)^2$
is finite also in the region of vanishing $\phi_b$~\cite{Hirvonen:2021xxx}
(c.f. also~\cite{Baacke:2003uw,Ekstedt:2021kyx,Gould:2021ccf}).

Because the bounce solution has to be solved numerically from
Eq.~\eqref{eq:thermalLOeom},
the exponents $\mathcal{B}_{0,1}$ are necessarily obtained by numerical integration.
Nevertheless, their gauge-independence can still be proven analytically.
The gauge-fixing parameter is absent at leading order since $Z$ and $V^\mathrm{eff}$ are both $\xi$-independent at this order, implying gauge-independence of both
the bounce solution $\phi_b$ and
exponent $\mathcal{B}_0$.
The gauge-invariance of $\mathcal{B}_1$
is not immediately obvious since $\Veff_{\rmii{NLO}}$ in Eq.~\eqref{eq:VNLO}
explicitly depends on 
$\xi$.
To proceed, we utilise
the Nielsen identities~\cite{Nielsen:1975fs,Fukuda:1975di}
in analogy to~\cite{Metaxas:1995ab}.
These identities have been discussed in
the context of finite temperature in
e.g.~\cite{Patel:2011th,Garny:2012cg,Espinosa:2016nld,Arunasalam:2021zrs}. 
The Nielsen identity (in $d$-dimensional Euclidean space)
\begin{align}
\label{eq:Nielsen:def}
  \xi\frac{\partial\Seff}{\partial\xi} &=
  - \int{\rm d}^{d}\vec{x}  
  \frac{\delta\Seff}{\delta\phi(x)}\,\mathcal{C}(x) 
  \;,
\end{align}
relates
the variation of
the effective action with the gauge parameter to
the corresponding Nielsen functional
\begin{align}
\mathcal{C}(x) &=
  \frac{ig}{2} 
  \int{\rm d}^{d}\vec{y}
  \Big\langle
  \chi(x) c(x) \bar{c}(y)
  \nn &\hphantom{{}\frac{ig}{2\sqrt{2}}\int{\rm d}^{d}\vec{y}}\times
  \Bigl[
    \partial_{i}B_{i}(y) + \sqrt{2} g\xi \phi\chi(y)
  \Bigr]
  \Big\rangle
\;,
\end{align}
which admits a derivative expansion \cite{Garny:2012cg}
\begin{align}
\label{eq:C:exp}
\mathcal{C}(x) &= 
    C(\phi)
  + D(\phi) (\partial_\mu \phi)^2
  - \partial_\mu \bigl( \tilde{D}(\phi) \partial_\mu \phi \bigr)
  + \ordo{\partial^4}
\;. 
\end{align}
Together with the expansion of the effective action~\eqref{eq:action3},
this yields the Nielsen identities for
the effective potential and field renormalization factor
\begin{align}
\label{eq:nielsen}
  \xi \frac{\partial}{\partial \xi} \Veff =&
    -  C \frac{\partial}{\partial \phi} \Veff
  \;,\\
\label{eq:nielsen:2}
  \xi \frac{\partial}{\partial \xi} Z =&
    - C \frac{\partial}{\partial \phi} Z
    - 2 Z \frac{\partial}{\partial \phi} C
    \nn &
    - 2 D  \frac{\partial}{\partial \phi} \Veff
    - 2 \tilde{D} \frac{\partial^2}{\partial \phi^2} \Veff
  \;.
\end{align}
To employ these relations, we expand them in powers of $g$
by first quoting the scaling of the Nielsen coefficients 
\begin{align}
\label{eq:CLO}
  C &= 
    C_{\rmii{LO}}
  + \mathcal{O}(g^{\frac{3}{2}} T)
  \;, \quad
C_{\rmii{LO}} = T \frac{\sqrt{\xi}}{16\pi} g
  \;, \\
  \label{eq:D}
  D &= \mathcal{O}(g^{-1} T^{-3})
  \;,\quad
  \tilde{D} = \mathcal{O}(g^{-1} T^{-2})
  \;.
\end{align}
These coefficients are computed at leading order 
in~\cite{Garny:2012cg,Hirvonen:2021xxx}.
We do not need the explicit expressions of $D, \tilde{D}$ because the terms on the second line of Eq.~\eqref{eq:nielsen:2}
appear at $\ordo{g^2}$, and are hence
suppressed relative to those on the first line at $\ordo{g}$.
The leading order $C_{\rmii{LO}} \sim g T$
is independent of the scalar background field $\phi$ at finite temperature.
An explicit counting in powers of $g$  
in the identities~\eqref{eq:nielsen} and \eqref{eq:nielsen:2} yields 
\begin{align}
\label{eq:highT-nielsen-1}
 \xi \frac{\partial}{\partial \xi} \Veff_{\rmii{NLO}} &=
  - C_{\rmii{LO}} \frac{\partial}{\partial \phi} \Veff_{\rmii{LO}}
  \;,\\
\label{eq:highT-nielsen-2}
\xi \frac{\partial}{\partial \xi} Z_{\rmii{NLO}} &= 
  -2  \frac{\partial}{\partial \phi} C_{\rmii{LO}}
  \;,
\end{align}
at
$\ordo{g^{4} T^4}$ in the first and
$\ordo{g^{ } T}$ in the second Nielsen identity.
A combination of the explicit expressions
\eqref{eq:VLO},
\eqref{eq:VNLO},
\eqref{eq:Z-NLO}, and
\eqref{eq:CLO}
readily verifies both identities.
In particular, the equality~\eqref{eq:highT-nielsen-2} holds since
the NLO field renormalisation is
$\xi$-independent and
at LO $C$ is $\phi$-independent.

Using the above Nielsen identities,
we demonstrate gauge independence of
$\mathcal{B}_1$:
\begin{align}
\label{eq:B1gaugedep}
  \xi \frac{\partial}{\partial \xi} \mathcal{B}_1 &=
  \xi \frac{\partial}{\partial \xi} \beta \int \diff^3 x \Bigl[
      \Veff_{\rmii{NLO}}(\phi_b)
    + \frac{1}{2}Z_{\rmii{NLO}}^{ }\left(\partial_\mu \phi_b\right)^2
  \Bigr]
  \nn
  & \stackrel{(A)}{=} \beta \int \diff{}^3 x \Bigl[
    - C_{\rmii{LO}}^{ } \frac{\partial}{\partial \phi} \Veff_{\rmii{LO}}(\phi_b)
    \Bigr]
  \nn & \stackrel{(B)}{=}
  - 
  C_{\rmii{LO}}^{ }\, \beta \int \diff{}^3 x \Bigl[ \square \phi_b\Bigr]
  \nn & \stackrel{(C)}{=}
  - 
  C_{\rmii{LO}}^{ }\, \beta \int \diff^2 S \cdot (\partial \phi_{b})
  \nn & \stackrel{(D)}{=}
  0
\;.
\end{align}
Step~$(A)$ uses the
Nielsen identity~\eqref{eq:highT-nielsen-1} and
$\xi$-independence of $Z_{\rmii{NLO}}$; 
$(B)$ applies the equation of motion~\eqref{eq:thermalLOeom} and
moves $C_{\rmii{LO}}$ outside the integrand due to its $\phi$-independence;
$(C)$ applies Gauss's theorem;
$(D)$ follows from the asymptotic behaviour of the bounce solution at the boundary,
\begin{align}
\label{eq:bubbleasymp}
  \square \phi_b &\sim \mu^2_{\rmii{eff}} \phi_b
  \;, \quad
  \phi_b(\infty)=0
  \;,\\
  \implies \phi_b(r) &\sim c \frac{e^{-\mu_{\rmii{eff}} r}}{r}
  \qquad
  \text{as } r \to \infty
  \;.
\end{align}
This completes our proof of gauge invariance of
the leading exponential of the nucleation rate. 
\begin{figure*}[t]
    \includegraphics[width=.5\textwidth]{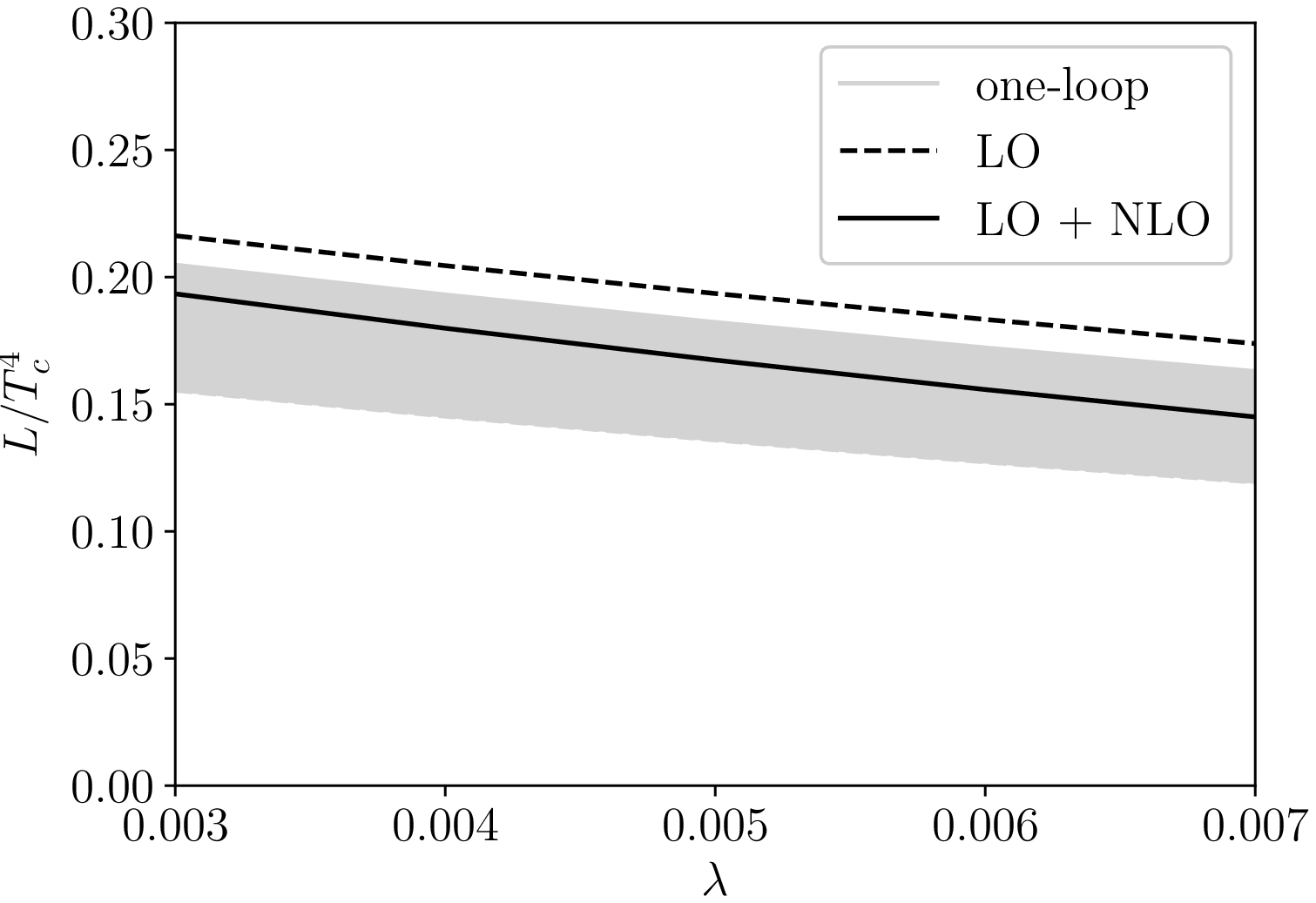}%
    \includegraphics[width=.5\textwidth]{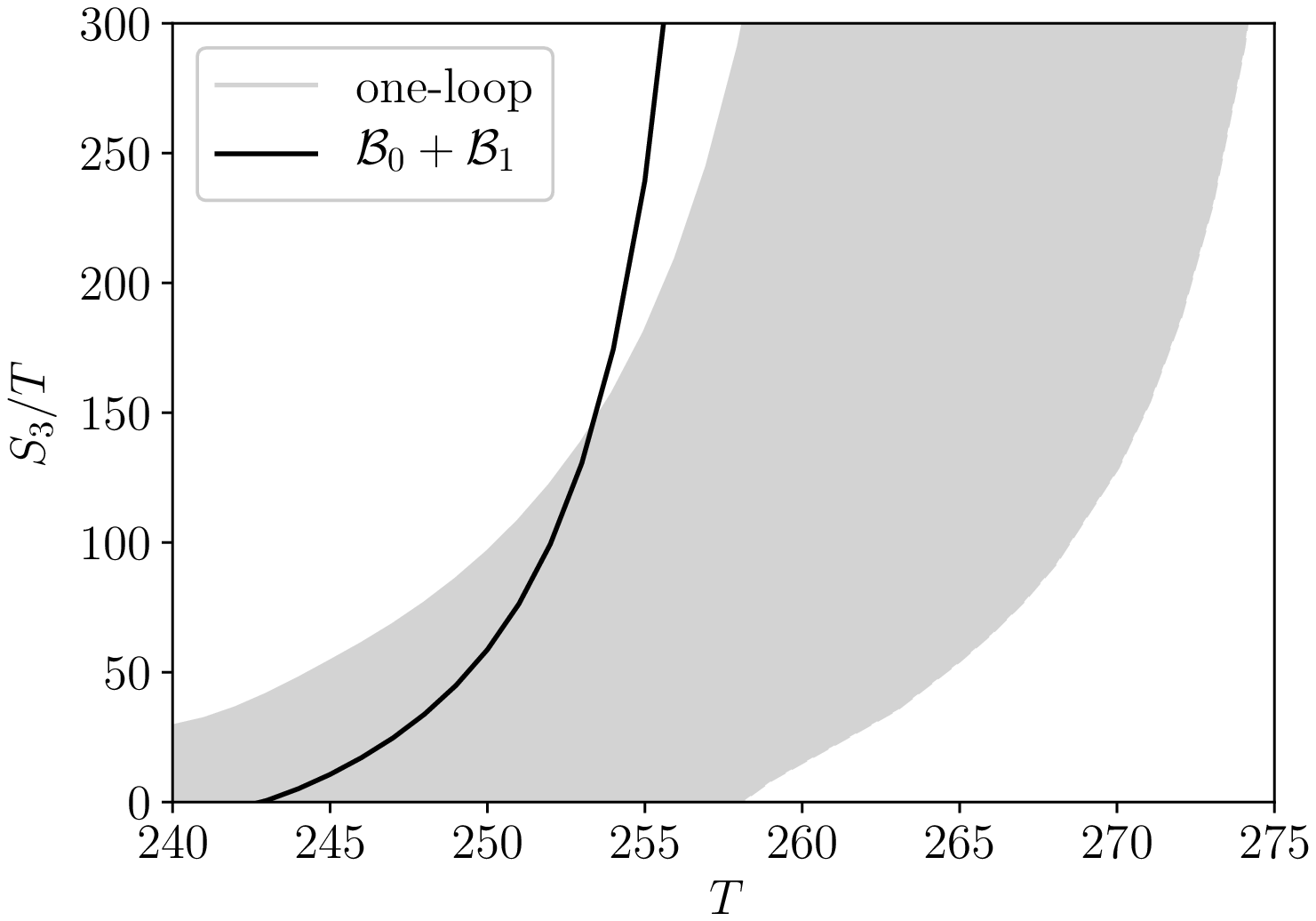}
  \caption{
  An illustration of
  a gauge independent, perturbative determination of the thermodynamics for a first order phase transition, with
  $g^2 = 0.42$ and
  $M = 100$, where $M$ and $T$ are in arbitrary units of mass.  
  Left:
  The dimensionless ratio
  $L/T^4_c$ 
  as function of $\lambda$ at
  LO (dashed line) and 
  NLO (solid line). 
  Right:
  The action
  $S_3/T = \mathcal{B}_0 + \mathcal{B}_1$
  as function of temperature at
  $\lambda = 0.005$, is shown by 
  a black solid curve.
  For comparison, both panels illustrate
  the conventional, inconsistent gauge dependent determination, wherein
  the light grey band shows the variation of
  the gauge-fixing parameter
  $\xi=(0,\dots,5)$.
  }
  \label{fig:numerics}
\end{figure*}

Figure~\ref{fig:numerics} illustrates the quantitative impact of applying our framework as compared to the conventional gauge-dependent approach. 
Fig.~\ref{fig:numerics} (left) gives
the $\lambda$-dependence of the strength of the transition, characterized by latent heat released in the transition, $L$,
scaled to $\Tc^4$.
Fig.~\ref{fig:numerics} (right) shows the $T$-dependence of $S_3/T$. 
We fix
$g^2 = 0.42$ and $\mu^2_{\rmii{eff}} = -M^2/2$, where
$M = 100$, and where
$M$ and $T$ are in arbitrary units of mass.  
Furthermore, we define $\lambda$ and $g$ at an initial
renormalisation scale $\Lambda_0 = 100$ and 
run to a scale $\Lambda = T$ using
one-loop renormalization group evolution~\cite{Hirvonen:2021xxx}.

The dashed (solid) line in Fig.~\ref{fig:numerics} (left)
corresponds to the gauge-invariant result for $L/\Tc^4$ at LO (LO$+$NLO).
For comparison, we also illustrate the conventional analysis based on direct minimization of the full one-loop effective potential (in the high-$T$ expansion).
The latter approach is gauge-dependent and inconsistent, as illustrated by the light grey band in which the gauge-fixing parameter ranges over $\xi=(0,\dots,5)$.
The qualitative behaviour -- the transition is stronger for smaller $\lambda$ --  agrees with the gauge-independent determination, but
the quantitative result is ambiguous. 

Fig.~\ref{fig:numerics} (right) shows the $T$-dependence of
$S_3/T = \mathcal{B}_0 + \mathcal{B}_1$  for fixed
$\lambda = 0.005$ as the 
black solid curve.
We have obtained this result using
{\tt FindBounce}~\cite{Guada:2020xnz} to find the bounce solution $\phi_b$ of
Eq.~\eqref{eq:thermalLOeom}.
For comparison,
the light grey band significantly varies in $S_3/T$
when varying $\xi = (0,\dots,5)$ as manifestation of 
an inconsistent computation that is often encountered in
the past literature (e.g.~\cite{Delaunay:2007wb,Caprini:2019egz}).
The latter result is based on computing
Eq.~\eqref{eq:rate} using Eq.~\eqref{eq:action3}, where the effective potential is directly computed at one-loop order in a fixed gauge. Therein, the bounce solution is found from the same gauge-dependent one-loop effective potential.
This leftover gauge-dependence immediately signals an inconsistency, as the computation of a physical quantity should contain no $\xi$-dependence.
For a discussion of other theoretical inconsistencies encountered in
the ``conventional'' approach, such as the appearance of an imaginary part in $S_3$ and double counting of contributions from the nucleating field, see~\cite{Croon:2020cgk,Gould:2021ccf}. The renormalization scale-dependence of the results in
Fig.~\ref{fig:numerics}, that can be used to monitor the accuracy of perturbation theory
(c.f.~\cite{Croon:2020cgk,Gould:2021oba}),
is further discussed in~\cite{Hirvonen:2021xxx}.

The framework presented in this letter provides a way to obtain
a gauge-independent, perturbative estimate of
the thermal nucleation rate in the presence of radiative barriers. 
While we have worked in the Abelian Higgs model along the lines of
previous literature~\cite{Metaxas:1995ab,Garny:2012cg}, our
framework readily generalises to more complicated gauge field theories,
with radiatively generated barrier,
and
our practical approach can facilitate
corresponding
model-building phenomenological studies.
Ultimately, one must assess the quantitative and qualitative reliability of
perturbative nucleation rate computations through comparison with non-perturbative calculations (see~\cite{Moore:2000jw,Moore:2001vf}).
A meaningful comparison requires a well-defined,
gauge-invariant perturbative computation,
which the framework presented herein provides.

\begin{acknowledgments}
{\em Acknowledgments.}---%
We acknowledge enlightening discussions with
Andreas Ekstedt,
Oliver Gould,
Joonas Hirvonen
and
Juuso {\"O}sterman and
specially thank Suntharan Arunasalam for his contributions during early stages of the project.
MJRM and TT are supported in part under National Science Foundation of China grant No.~19Z103010239.
PS has been supported
by the European Research Council, grant no.~725369, and
by the Academy of Finland, grant no.~1322507.
TT thanks Aleksi Vuorinen and Helsinki Institute of Physics for their hospitality and support during part of this work.
\end{acknowledgments}

\appendix
\allowdisplaybreaks
\bibliographystyle{apsrev4-1}
\bibliography{references}

\end{document}